# Social capital and resilience make an employee cooperate for coronavirus measures and lower his/her turnover intention

Keisuke Kokubun[1*], Yoshiaki Ino[2], and Kazuyoshi Ishimura[2]

1 Economic Research Institute, Japan Society for the Promotion of Machine Industry
2 IEWRI Japan Co., Ltd.

# Abstract

**Purpose** – An important theme is how to maximize the cooperation of employees when dealing with crisis measures taken by the company. Therefore, the purpose of this paper is to find out what kind of employees cooperated with the company's measures in the current corona (COVID-19) crisis, and what effect the cooperation had to these employees/companies to get hints for preparing for the next crisis.

**Design/methodology/approach** – Pass analysis was carried out using awareness data obtained from a questionnaire survey conducted on 2,973 employees of Japanese companies in China.

**Findings** – The results showed that employees with higher social capital and resilience were more supportive of the company's measures against corona and that employees who were more supportive of corona measures were less likely to have turnover intention. However, regarding fatigue and anxiety about corona felt by employees, it was shown that it not only works to support cooperation in corona countermeasures but also enhances the intention to leave.

**Research limitations/implications** – This study used self-report data from individual respondents, which may have resulted in common method bias, although we confirmed that the impact was not large. Future research might consider including supervisor-rated scales to strengthen the study design and reduce common method bias.

**Practical implications** – It is important for employees to be aware of the crisis and to fear it properly. But more than that, it should be possible for the company to help employees stay resilient, build good relationships with them, and increase the social capital to make them support crisis measurement of the company most effectively while keeping their turnover intention low.

**Originality/value** – This research is one of the first papers to show how effective human resource management daily is in crisis management, focusing on measures against corona.

Keywords: Coronavirus (COVID-19), Resilience, Social capital, Anxiety, Fatigue, Countermeasures, Cooperation, Turnover intention

**Introduction**

In the new coronavirus (COVID-19) peril, Japanese companies in China were required to deal with employee hygiene and attendance management. Among these, what kind of company was able to get employees to work on corona measures smoothly? Also, how was such cooperation related to the employee's turnover intention, which is an important issue in personnel measures? In this paper, we would like to consider this problem by conducting an analysis using data obtained from a questionnaire survey recently conducted on employees of Japanese companies in China.

The first possible factor that determines the cooperation of employees in dealing with a company's crisis is how they have accepted their work and company. Employees who take a positive attitude about work and the company daily may have been more supportive of the company's measures against corona. Also, as a side effect, if the measures of the organization are successful through cooperation, cohesion among employees may have been strengthened and the willingness to leave may have declined. In China, where a Confucian tradition has a strong influence, the relationship of reciprocity (employees who feel the support of the organization to act in return for the organization) is still common today. Previous studies have argued that it still has a significant impact on the human resources policies of modern enterprises in China (e.g., Warner, 2010).

The next thing to consider is how responsive the employees were to the crisis. Many previous studies have revealed that the crisis response capabilities of employees play an important role in recovering from a crisis, using keywords such as resilience and coping as a clue (e.g., Aldrich, 2012). However, most of them are studies that deal with the ability to heal and recover the feelings that have been hurt by a crisis such as a natural disaster. What impact does employee resilience or coping have on the company's measurements against the crisis? In a continuous and difficult-to-see crisis such as this coronavirus peril, there is a large part that cannot be overcome by individual tolerance alone, so it is meant to clarify how it is connected to the organization's attitude to crisis countermeasures.

The last thing to consider is how anxious employees were about the coronavirus peril. The magnitude of anxiety will depend not only on the personality of the individual, such as whether or not he/she is prone to anxiety but also on what kind of health condition he/she was in at work. Employees who are well-managed in good physical and mental health may not have had much anxiety during this crisis. On the other hand, employees who are tired due to poor health management may have become more and more anxious during this crisis and may have been more eager to seek corona measures.

Furthermore, if the employee's efforts to combat corona are short-term outcomes, the long-term outcomes the organization should aim for in human resource management, including lower employee turnover intentions, are also need to be discussed. By doing so in the current research, we believe that it will give us a hint to take measures during a crisis without sacrificing employee loyalty to the organization and, if possible, increasing it further.

**Theoretical framework**
It is important to focus on resilience when considering how to respond to a crisis. Resilience refers to the ability of people, society, and organizations to control damage and increase sustainability in adversity (Serfilippi and Ramnath, 2018). To measure a person's/organization's resilience, it is necessary to look at how the person/organization was exposed to adversity and how it responded (Sutcliffe and Vogus, 2003). Furthermore, the evaluation of the response will change depending on whether it was a temporary measure for overcoming adversity or a long-term effect. To that end, resilience-based research needs to clearly define the adversity faced by the person/organization, the response it has taken, and the short- and long-term outcomes (Rao and Greve, 2018). In this paper, we will discuss corona erosion as the adversity faced by employees/organizations, corona countermeasures represented by hygiene/ attendance management as the measures taken by the organizations, employees' cooperation in corona countermeasures as the short-term result, and the employees' intention to leave as the long-term result. In other words, we consider not only how much employees were able to be involved in the measures taken by the organization during the crisis, but also how much they were able to lower their intention to leave the job as the result of a series of measures against corona. This is because, if the tactical approach to a crisis can influence employee loyalty to the organization, simply overcoming the immediate crisis is unlikely to be a complete success. As a factor that influences these outcomes, we adopt social capital, resilience, and anxiety about the corona for the reasons mentioned at the beginning. Furthermore, we would like to briefly confirm the validity of the analytical model by picking up the voices in the open-ended column provided in the survey form.

**Determinants of cooperation intentions for corona countermeasures**

So how can we get our members to help the organization deal with corona? If there are differences in the attitudes of employees' cooperation between organizations, it can be said in a nutshell that they are due to the accumulation of daily personnel measures. In considering this problem, we would like to refer to social capital, a characteristic of social organizations such as trust, norms, and networks that can improve social efficiency by activating people's coordinated actions (Putnam et al., 1994). It should be recalled here that many organizations routinely take various measures to increase the social capital of their employees. This is because members are said to unite and exchange information to achieve their goals in high social capital organizations (Leana and Van Buren, 1999). In line with social exchange theory-based literature on employee-employer relationships (Coyle-Shapiro and Shore, 2007; Cropanzano and Mitchell, 2005), a previous study confirmed that social capital can contribute to employees' sense of obligation to pay back the cooperative and collaborative working atmosphere (Parzefall and Kuppelwieser, 2012). In particular, concerning disasters such as the corona eruption, which was difficult to predict in advance, it is thought that employees were required to have more understanding and cooperation than what can be captured by written rules and manuals. Therefore, it is considered that the presence or absence of social capital influenced the size of cooperation for the measures against corona. From the above, the following hypotheses are derived.

H1: Employees with higher social capital were more supportive of the organization's measures against corona

Previous studies have shown that highly resilient employees can respond appropriately to organizational problems (Avey et al., 2009) and tend to be more flexible in accepting and cooperating with organizational changes (Shin et al., 2012; Wanberg and Banas, 2000). Therefore, it is expected that employees with higher resilience were more supportive of the organization's measures against corona, and the following hypothesis is derived.

H2: Employees with higher resilience were more supportive of the organization's measures against corona

However, no matter how large the social capital and resilience are, if the corona isn't something to worry about, the need for corona countermeasures will not be felt, and as a result, cooperation with the organization will be limited. Anxiety can sometimes be a motivation to take specific action by raising awareness of dangers to avoid and problems to address (Strack et al., 2017). Therefore, if there is a great deal of anxiety about the corona, it is considered that they are willing to cooperate with corona countermeasures, and the following hypothesis is derived.

H3: Employees who were more concerned about corona were more supportive of the organization's measures against corona

**Determinants of turnover intention**

Many researchers have pointed out that infectious diseases have historically caused anxiety, suspicion, and panic, weakening trust in the policies of the organization, or weakening the cohesion of the organization (Edelstein, 1988; Peckham, 2015; Picou et al., 2004). Relatedly, Erikson (1994) argues that if the disaster is artificial, it weakens the cohesion of the organization. Such a phenomenon may occur even when the members' feelings leave the organization when the organization cannot effectively take measures against the disaster and cannot prevent the damage that would otherwise be prevented. Therefore, the following hypothesis is derived.

H4: Employees with greater anxiety were more likely to leave

But at the same time, Erikson (1994) also mentions the potential of natural disasters to strengthen organizational cohesion. It can be assumed that if the members cooperate with the disaster countermeasures of the organization and the response is successful, the trust in the organization increases. These arguments may imply that disasters normally act by damaging members' loyalty to the organization by causing anxiety, but if an organization can make its members cooperate to the measures and respond well to disasters, it is not impossible to increase their loyalty. Therefore, the following hypothesis is derived.

H5: Employees who were more supportive of the organization's measures against corona were less likely to leave

Previous studies have shown that social capital has the effect of reducing turnover rates (Shaw et al., 2005). Therefore, the following hypothesis is derived.

H6: Employees with higher social capital were less likely to leave

**Relationship between social capital and resilience**

Lengnick-Hall (2011) reviews a series of previous studies and argues that practicing HR policies that create social capital and solidarity creates resilience for employees and organizations. Therefore, the following hypothesis will be established.

H7: Employees with higher social capital had higher resilience

**Relationship between resilience and anxiety**

It is considered that those with high resilience were less anxious about corona because of their confidence that they could cope well. Empirical findings indicate a negative correlation between resilience and anxiety (Hjemdal et al., 2011; Shi et al., 2015). Therefore, the following hypothesis is derived.

H8: Employees with higher resilience were less concerned about the corona

**Relationship between fatigue and anxiety**

In a workplace where health care is not routinely maintained, it is considered that employees became tired and became more and more concerned that the company would not take adequate measures against the corona. Correlation between fatigue and anxiety has also been shown in previous studies (Jiang et al., 2003; Kokubun et al., 2018). Therefore, the following hypothesis is derived.

H9: More tired employees were more concerned about the corona

**Relationship between social capital and fatigue**

The social capital variable was negatively correlated with fatigue in the previous studies (Lim et al., 2016; Miller et al., 2006). Therefore, the following hypothesis is derived.

H10: Employees with greater fatigue had a lower social capital

**Relationship between fatigue and intention to leave**

Accumulation of fatigue will increase the desire for a less fatiguing workplace. Previous studies have shown that fatigue and stress increase the willingness to leave the job (De Croon et al., 2004; Lee and Jang, 2020). Therefore, the following hypothesis is derived.

H11: More tired Employees were more likely to leave

From the above, the 11 hypotheses shown in Fig. 1 will be tested in this paper.

Figure 1 Hypothesis tested in this paper

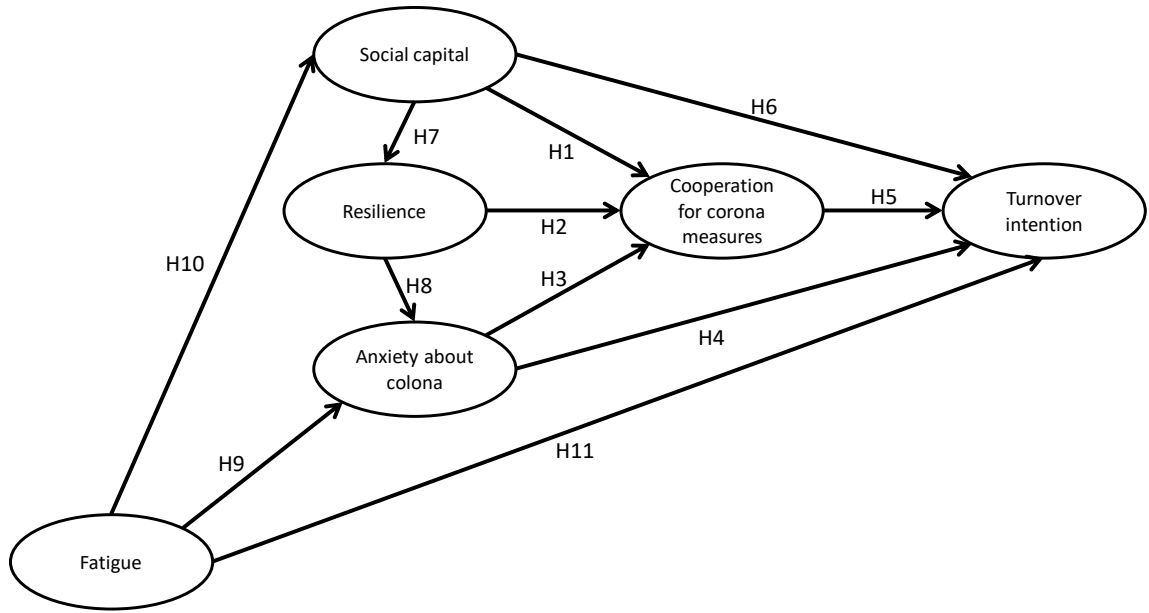

## Research method

The questionnaire consisted of attribute items such as gender and residential area, as well as 40 question items based on the 5-point method. Of these, the social capital was created by modifying and adding the organizational support items of Eisenberger et al. (1986) to organizational commitment items of Kokubun (2018) which was created referring to Mowday et al. (1982). In other words, we define social capital as the reciprocal state of organizational support and employee commitment. Putnam et al. (1994) defined that social capital consists of three elements: trust, reciprocity, and social networks. However, this paper, which focuses on relationships within an organization, defines social capital by focusing on trust (organizational support) and reciprocity (employee commitment).

Besides, fatigue was composed of the three items of Kokubun (2017; 2018) with the addition of independently developed two items: "My work always keeps me from feeling" and "I sometimes feel frustrated while working". This is because to see the relationship between the anxiety about the corona and the mental state, we thought that not only physical but also mental fatigue was more appropriate in the analytical model. Regarding resilience, it was created by referring to the resilience scale of Saito and Okayasu (2010) developed referring to Oshio et al. (2002), Wagnild et al. (1993), Friborg et al. (2003), Connor et al. (2003), and disaster self-efficacy scale of Motoyoshi (2017) developed referring to Nypaver (2011). Other corona-related items were independently developed referring to various media and articles.

From February 15th to May 31st, 2020, this questionnaire, after being translated into Chinese by the back-translation method, was distributed to more than 10,000 people at 32 companies in the eastern and southern areas, and 6,673 people answered. However, this analysis uses data from 2,973 people from 26 companies who answered all the questions. Most of the target people are manufacturing employees (94.7%) and the rest are design, finance, consulting, etc.

**Analysis and results**

All statistical analyses have been performed using IBM SPSS Statistics/AMOS Version 26 (IBM Corp., Armonk, NY, USA). As a result of factor analysis by varimax rotation, six factors as shown in Table 1 were extracted. The factor names of social capital (composed of 14 items), resilience (composed of 10 items), cooperation for corona measures (composed of 5 items), fatigue (composed of 5 items), anxiety about corona (composed of 4 items), and turnover intention (composed of 2 items) were assigned.

Before proceeding to the main analyses, Harman's single-factor analysis was used to check if the variance in the data could be largely attributed to a single factor, while the confirmatory factor analysis (CFA) was used to test if the factors were related to the measures. First, the factor analysis indicated that only 36.3 percent of the variance could be explained by a single factor, which was <50 percent. Thus, it was established that the data did not suffer from the common method variance (Chang et al., 2010). Next, for CFA, the model fit was evaluated by examining the indices recommended by Hu and Bentler (1999) and it was shown that the 6-factor model shown in Table 1 fits better than the 1-factor model that added 6 variables (available upon request for more details).

Table 2 shows the results of descriptive statistics and correlation analysis between variables. For each variable, 1 to 5 points are assigned to the individual response items of the 5-scale method, and the average is calculated for easy comparison. The highest score was 4.639 for cooperation for corona measures, followed by 4.332 for resilience and 3.922 for social capital. On the contrary, the lowest was 1.812 for turnover intention, followed by 2.508 for fatigue and 3.630 for anxiety about the corona. It can be seen that the overall positive awareness is high and the negative awareness is low. However, looking at the standard deviation, the former is 0.707 to 0.940, while the latter is 1.121 to 1.169, indicating that the latter has more variation. Therefore, it should be noted that the negative consciousness, especially the turnover intention, is not so high on average, but the difference between employees is relatively large. Regarding age, 1 to 4 points were assigned to the options of "under 30", "30-39", "40-49", and "50 or older". Similarly, concerning the length of service, 1 to 4 points were assigned to the options "less than 1 year", "1 year to less than 3 years", "3 years to less than 5 years", and "5 years or more".

Tables 3, 4, and 5 are descriptive statistics by gender, region, and position, respectively. In terms of gender, males were significantly lower than females in social capital, cooperation with corona measures, and anxiety about coronas and males were significantly higher than females in intention to leave. By region, the eastern region was significantly higher in terms of social capital, and significantly lower in fatigue, anxiety about the corona, and turnover intention, than the southern region. By position, managers were significantly higher than non-managers for cooperation with social capital, resilience, and measures against corona and managers were significantly lower than non-managers for fatigue, anxiety about the corona, and turnover intention. Therefore, the following analysis will be performed after incorporating gender, region, and position into the variables.

Fig. 2 shows the result of the path analysis. Gender/regional/positional paths and covariance between variables are omitted. However, in Table 5, the direct effect, the indirect effect, and the total effect of the path coefficient are listed for all variables including the variables omitted in Fig. 2. First, it was shown that higher social capital and resilience would lead to greater cooperation in combating corona and supported H1 and H2. H3 was also supported by showing that greater anxiety about corona increases cooperation with corona countermeasures. Furthermore, H4, H5, and H6 were supported as it was shown that the anxiety about corona heightened the intention to leave the job, while the cooperation to the measures against corona and the social capital lowered the intention to leave the job. Besides, the relationship between higher social capital and higher resilience was shown, supporting H7. H9, H10, and H11 were also endorsed, as fatigue was shown to increase anxiety about the corona, reduce social capital, and increase willingness to leave. The only unexpected thing was the relationship between resilience and anxiety about the corona. In other words, it was the opposite of the expectation that an increase in resilience would increase anxiety about the corona, resulting in the rejection of H8. By the way, Table 3 shows that although the resilience (0.580) is the largest in terms of the overall effect on cooperation in countermeasures against corona, social capital (0.507) and anxiety about corona (0.172) are also relatively large. Regarding the overall effect on turnover intention, fatigue (0.410) was the largest, followed by social capital (-0.229).

Thus, 10 out of 11 hypotheses were supported. Only the rejected H8 resulted in the opposite of the assumption, that is, the higher the resilience, the greater the anxiety about the corona.

Table 1 Results of factor analysis

| item | Social capital | Resilience | Cooperation for corona measures | fatigue | Anxiety about corona | Turnover intention |
| --- | --- | --- | --- | --- | --- | --- |
| I like working in a company | **0.807** | 0.173 | 0.099 | -0.153 | 0.023 | -0.106 |
| I am generally satisfied with my current work | **0.773** | 0.144 | 0.103 | -0.169 | -0.011 | -0.062 |
| Work at the company is fun | **0.748** | 0.160 | 0.075 | -0.210 | -0.026 | -0.020 |
| The company cares about its employees | **0.741** | 0.194 | 0.131 | -0.156 | -0.035 | 0.001 |
| I find my job very rewarding | **0.739** | 0.167 | 0.038 | -0.146 | -0.026 | -0.006 |
| The company trusts its employees | **0.735** | 0.209 | 0.163 | -0.126 | 0.014 | -0.033 |
| I am proud of my work | **0.716** | 0.144 | 0.017 | -0.153 | -0.048 | 0.018 |
| I want to work hard at this company | **0.712** | 0.231 | 0.214 | -0.036 | 0.020 | -0.233 |
| The company treats its employees fairly | **0.708** | 0.178 | 0.063 | -0.185 | -0.075 | 0.045 |
| I want to work for this company forever | **0.701** | 0.240 | 0.192 | -0.034 | 0.019 | -0.255 |
| I want to continue my current work | **0.698** | 0.211 | 0.167 | -0.076 | 0.033 | -0.177 |
| The company is reliable | **0.689** | 0.244 | 0.289 | -0.071 | 0.049 | -0.144 |
| I want to contribute to the growth of the company | **0.669** | 0.263 | 0.234 | -0.027 | 0.011 | -0.193 |
| Satisfaction with work content | **0.545** | 0.290 | 0.113 | -0.232 | -0.016 | -0.009 |
| I think I can handle various things well even in a mess | 0.207 | **0.843** | 0.119 | -0.034 | 0.025 | -0.015 |
| I think I can flexibly respond to various things when an emergency occurs | 0.197 | **0.819** | 0.128 | -0.023 | 0.034 | -0.010 |
| I think I can act calmly even in an emergency | 0.216 | **0.804** | 0.164 | -0.055 | 0.034 | -0.017 |
| I think I can stay relatively calm even in the chaos | 0.207 | **0.769** | 0.150 | -0.050 | 0.029 | -0.018 |
| I think I can overcome the pain and tragedy | 0.220 | **0.727** | 0.208 | -0.048 | 0.024 | -0.029 |
| I am confident that I will manage to live even if I encounter difficulties | 0.249 | **0.688** | 0.304 | -0.058 | 0.074 | -0.090 |
| I think I have the power to achieve my goals | 0.282 | **0.614** | 0.244 | -0.092 | 0.068 | -0.027 |
| I think that if I make an effort, I can solve anything by myself | 0.256 | **0.599** | 0.258 | -0.060 | 0.151 | -0.064 |
| Even if something unpleasant happens, I often think that my present experience should be good for the future | 0.278 | **0.550** | 0.324 | -0.064 | 0.042 | -0.095 |
| No matter how difficult the situation is, I will not give up | 0.355 | **0.524** | 0.361 | -0.102 | 0.104 | -0.134 |
| I would like to cooperate with the hygiene management of the company to prevent new coronavirus infection | 0.188 | 0.332 | **0.783** | -0.006 | 0.149 | -0.077 |
| I would like to cooperate with the instructions of the company related to work attendance to prevent the new coronavirus infection | 0.196 | 0.355 | ***0.780*** | 0.021 | 0.135 | -0.073 |
| The company gives its employees information and measures regarding the new coronavirus | 0.212 | 0.328 | ***0.759*** | -0.004 | 0.133 | -0.038 |
| The company has adequate hygiene measures against the new coronavirus | 0.314 | 0.334 | ***0.620*** | -0.057 | 0.100 | -0.040 |
| I personally take measures against the new coronavirus infection | 0.172 | 0.348 | ***0.619*** | 0.015 | 0.102 | -0.023 |
| I am often tired and tired | -0.169 | -0.043 | 0.010 | ***0.829*** | 0.160 | 0.033 |
| I am exhausted when I finish my work | -0.157 | -0.042 | 0.060 | ***0.731*** | 0.183 | 0.024 |
| My work always keeps me from feeling | -0.244 | -0.072 | -0.027 | ***0.715*** | 0.136 | 0.135 |
| I sometimes feel frustrated while working | -0.242 | -0.083 | -0.048 | ***0.654*** | 0.108 | 0.180 |
| I'm tired since I woke up in the morning | -0.170 | -0.066 | -0.034 | ***0.619*** | 0.144 | 0.140 |
| I'm worried about the new coronavirus | 0.045 | 0.070 | 0.128 | 0.090 | ***0.707*** | 0.002 |
| Feelings are blocked due to anxiety about the new coronavirus | -0.056 | 0.024 | -0.030 | 0.258 | ***0.652*** | 0.133 |
| My co-workers are worried about getting the new coronavirus at the company | -0.085 | 0.070 | 0.080 | 0.206 | ***0.607*** | 0.073 |
| I'm feeling financially uncertain because of the new coronavirus | 0.035 | 0.116 | 0.245 | 0.110 | ***0.548*** | -0.016 |
| Within a half year, I will quit my current job | -0.227 | -0.084 | -0.106 | 0.265 | 0.121 | ***0.816*** |
| After half a year, I will quit my current job | -0.213 | -0.084 | -0.073 | 0.277 | 0.114 | ***0.806*** |

Note) If the factor load is 0.4 or more, italic and bold type.

Table 2 Results of descriptive statistics and correlation analysis

| | | average | standard deviation | α | 1 | 2 | 3 | 4 | 5 | 6 | 7 |
|---|---|---|---|---|---|---|---|---|---|---|---|
| 1 | Social capital | 3.922 | 0.940 | 0.952 | | | | | | | |
| 2 | Resilience | 4.332 | 0.778 | 0.945 | 0.564** | | | | | | |
| 3 | Cooperation for corona measures | 4.639 | 0.707 | 0.929 | 0.471** | 0.659** | | | | | |
| 4 | Fatigue | 2.508 | 1.169 | 0.872 | -.394** | -.198** | -0.082** | | | | |
| 5 | Anxiety about corona | 3.630 | 1.121 | 0.753 | -0.028 | 0.144** | 0.250** | 0.353** | | | |
| 6 | Turnover intention | 1.812 | 1.126 | 0.923 | -.376** | -.235** | -0.209** | 0.424** | 0.202** | | |
| 7 | Age | 1.960 | 0.785 | | 0.154** | 0.130** | 0.082** | -.138** | -.062** | -.184** | |
| 8 | Length of service | 2.758 | 1.134 | | 0.014 | 0.034 | 0.083** | 0.024 | -0.001 | -.113** | 0.421** |

**. Significant at 1% level.  *. Significant at 5% level.  n = 2,973

Table 3 Descriptive statistics by gender

|  | Male (n=1,122) | | Female (n=1,851) | | t-value | p |
|---|---|---|---|---|---|---|
|  | Average value | standard deviation | Average value | standard deviation | | |
| Social capital | 3.865 | 0.994 | 3.956 | 0.904 | -2.555 | * |
| Resilience | 4.320 | 0.834 | 4.340 | 0.743 | -0.681 | |
| Cooperation for corona measures | 4.560 | 0.782 | 4.686 | 0.653 | -4.747 | ** |
| Fatigue | 2.527 | 1.165 | 2.496 | 1.172 | 0.690 | |
| Anxiety about corona | 3.516 | 1.155 | 3.699 | 1.094 | -4.335 | ** |
| Turnover intention | 1.956 | 1.175 | 1.724 | 1.086 | 5.480 | ** |
| Age | 1.860 | 0.811 | 2.021 | 0.763 | -5.444 | ** |
| Length of service | 2.656 | 1.119 | 2.821 | 1.139 | -3.846 | ** |

**. Significant at 1% level.  *.Significant at 5% level.  n = 2,973

Table 4 Descriptive statistics by region

|  | Eastern region (n=807) | | Southern region (n=2,166) | | t-value | p |
|---|---|---|---|---|---|---|
|  | Average value | standard deviation | Average value | standard deviation | | |
| Social capital | 4.038 | 0.876 | 3.878 | 0.959 | 4.117 | ** |
| Resilience | 4.330 | 0.729 | 4.333 | 0.796 | -0.074 | |
| Cooperation for corona measures | 4.658 | 0.674 | 4.632 | 0.719 | 0.897 | |
| Fatigue | 2.418 | 1.099 | 2.542 | 1.192 | -2.574 | * |
| Anxiety about corona | 3.170 | 1.120 | 3.801 | 1.072 | -14.098 | ** |
| Turnover intention | 1.575 | 0.974 | 1.900 | 1.166 | -7.051 | ** |
| Age | 1.979 | 0.824 | 1.953 | 0.771 | 0.789 | |
| Length of service | 2.629 | 1.231 | 2.807 | 1.092 | -3.794 | ** |
|  | n | % | n | % | χ2 | |
| Male | 249 | 30.9 | 873 | 40.3 | 22.344 | ** |
| Female | 558 | 69.1 | 1293 | 59.7 | | |

**. Significant at 1% level.  *.Significant at 5% level.  n = 2,973

Table 5 Descriptive statistics by position

|  | Management (n=150) | | Non-manager (n=2,823) | | t-value | p |
|---|---|---|---|---|---|---|
|  | Average value | standard deviation | Average value | standard deviation |  |  |
| Social capital | 4.222 | 0.823 | 3.906 | 0.943 | 4.025 | ** |
| Resilience | 4.535 | 0.604 | 4.321 | 0.785 | 3.288 | ** |
| Cooperation for corona measures | 4.785 | 0.482 | 4.631 | 0.716 | 2.609 | ** |
| Fatigue | 2.132 | 1.032 | 2.528 | 1.172 | -4.053 | ** |
| Anxiety about corona | 3.060 | 1.112 | 3.660 | 1.113 | -6.436 | ** |
| Turnover intention | 1.423 | 0.876 | 1.832 | 1.134 | -4.346 | ** |
| Age | 2.580 | 0.744 | 1.927 | 0.774 | 10.083 | ** |
| Length of service | 3.580 | 0.884 | 2.715 | 1.129 | 9.232 | ** |
|  | n | % | n | % | $\chi^2$ |  |
| Male | 101 | 67.3 | 1021 | 36.2 | 58.879 | ** |
| Female | 49 | 32.7 | 1802 | 63.8 |  |  |
| Eastern region | 62 | 41.3 | 745 | 26.4 | 16.082 | ** |
| Southern region | 88 | 58.7 | 2078 | 73.6 |  |  |

\*\*. Significant at 1% level.　\*.Significant at 5% level.　n = 2,973

Figure 2　Results of path analysis

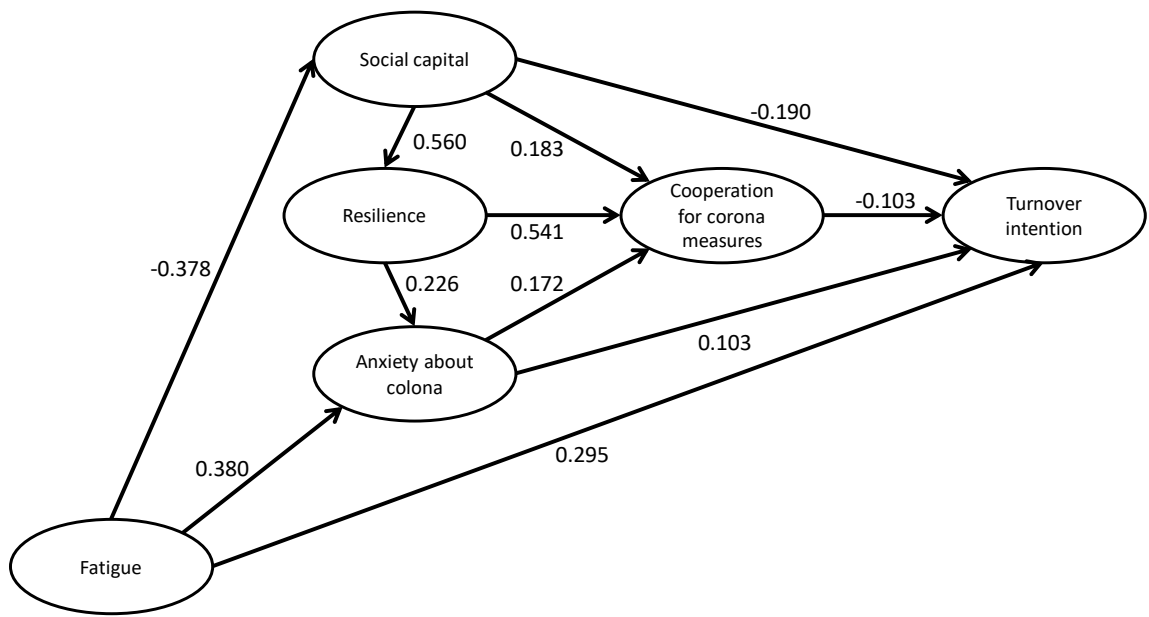

All passes are significant at the 1 % level. Goodness-of-fit indices: χ2 = 23.071; df = 15; root mean square error of approximation (RMSEA) = 0.013; probability of close fit (PCLOSE) = 1.000; goodness of fit index (GFI) = 0.999; adjusted goodness of fit index (AGFI) = 0.994; normed fit index (NFI) = 0.996; comparative fit index (CFI) = 0.999. n = 2,973.

Table 6 Path coefficient

| | | | Direct effect | Indirect effect | Overall effect |
|---|---|---|---|---|---|
| Management | ---> | Fatigue | -0.061 | | -0.061 |
| Age | ---> | Fatigue | -0.172 | | -0.172 |
| Length of service | ---> | Fatigue | 0.106 | | 0.106 |
| Eastern region | ---> | Social capital | 0.056 | | 0.056 |
| Fatigue | ---> | Social capital | -0.378 | | -0.378 |
| Age | ---> | Social capital | 0.101 | 0.065 | 0.166 |
| Management | ---> | Social capital | | 0.023 | 0.023 |
| Length of service | ---> | Social capital | | -0.040 | -0.040 |
| Social capital | ---> | Resilience | 0.560 | | 0.560 |
| Eastern region | ---> | Resilience | -0.044 | 0.031 | -0.013 |
| Age | ---> | Resilience | 0.045 | 0.093 | 0.138 |
| Management | ---> | Resilience | | 0.013 | 0.013 |
| Length of service | ---> | Resilience | | -0.022 | -0.022 |
| Fatigue | ---> | Resilience | | -0.212 | |
| Gender | ---> | Anxiety about corona | 0.096 | | 0.096 |
| Fatigue | ---> | Anxiety about corona | 0.380 | -0.048 | 0.332 |
| Resilience | ---> | Anxiety about corona | 0.226 | | 0.226 |
| Eastern region | ---> | Anxiety about corona | -0.236 | -0.003 | -0.239 |
| Management | ---> | Anxiety about corona | -0.066 | -0.020 | -0.086 |
| Age | ---> | Anxiety about corona | -0.034 | -0.034 | -0.068 |
| Length of service | ---> | Anxiety about corona | | 0.035 | 0.035 |
| Social capital | ---> | Anxiety about corona | | 0.127 | 0.127 |
| Social capital | ---> | Cooperation for corona measures | 0.183 | 0.325 | 0.507 |
| Resilience | ---> | Cooperation for corona measures | 0.541 | 0.039 | 0.580 |
| Anxiety about corona | ---> | Cooperation for corona measures | 0.172 | | 0.172 |
| Eastern region | ---> | Cooperation for corona measures | 0.050 | -0.038 | 0.012 |
| Gender | ---> | Cooperation for corona measures | 0.053 | 0.017 | 0.069 |
| Fatigue | ---> | Cooperation for corona measures | 0.031 | -0.126 | -0.095 |
| Age | ---> | Cooperation for corona measures | -0.041 | 0.088 | 0.047 |
| Length of service | ---> | Cooperation for corona measures | 0.078 | -0.010 | 0.068 |
| Management | ---> | Cooperation for corona measures | | -0.006 | -0.006 |
| Cooperation for corona measures | ---> | Turnover intention | -0.103 | | -0.103 |
| Gender | ---> | Turnover intention | -0.069 | 0.003 | -0.066 |
| Fatigue | ---> | Turnover intention | 0.295 | 0.116 | 0.410 |
| Social capital | ---> | Turnover intention | -0.190 | -0.039 | -0.229 |
| Anxiety about corona | ---> | Turnover intention | 0.103 | -0.018 | 0.085 |
| Eastern region | ---> | Turnover intention | -0.072 | -0.037 | -0.108 |
| Length of service | ---> | Turnover intention | -0.085 | 0.036 | -0.050 |
| Age | ---> | Turnover intention | -0.055 | -0.094 | -0.149 |
| Management | ---> | Turnover intention | | -0.031 | -0.031 |
| Resilience | ---> | Turnover intention | | -0.036 | -0.036 |

| Covariance | | | |
|---|---|---|---|
| Gender | <--> | Eastern region | 0.085 |
| Management | <--> | Length of service | 0.167 |
| Management | <--> | Age | 0.181 |
| Length of service | <--> | Age | 0.422 |
| Management | <--> | Gender | -0.141 |
| Length of service | <--> | Gender | 0.070 |
| Age | <--> | Gender | 0.098 |
| Management | <--> | Eastern region | 0.071 |
| Length of service | <--> | Eastern region | -0.076 |

Note) Numbers are standardized estimates. All are significant at the 5% level.

Free answer analysis

The participants of the survey were asked to comment freely under the description "What suggestions do you have for the current situation?" Regarding the measurements against coronal bruise, there were: "Always open windows and improve ventilation" "Work to prevent coronavirus" "Want to distribute mask regularly" "Share latest information" "Coronavirus has had a huge impact on our employees, so we would like to ask our employees to assist after the convergence".

     However, more common were the following comments that argued for the importance of unity: "Let's work together to fight the virus, overcome the difficulties, and

do our best together!" "Continuing attention to the current situation of the coronavirus, all employees will work together" "I hope that the company can overcome this difficulty" "I will do my best to move forward together with the company" "I will do my best to overcome the difficulties together" "Let's unite and survive this coronal disaster together."

Besides, there were noticeable comments for after the disaster: "Difficulty is temporary, so I hope that the company will develop more and more" "The crisis brought by Corona Erosion is also an opportunity. We can make better use of our strengths" "Both coronavirus prevention and production activities must not be neglected" "Encourage the production and recover the loss of corona damage" "The difficulty of corona damage is temporary, so I hope that the company will recover quickly and production will be normal!"

This fact is compatible with the model in this paper, in which a good relationship between the company and employees daily leads to cooperation with employees for measures against corona (a short-term result) and suppression of their intention to leave the company (a long-term result).

**Discussion**

First, it was revealed that employees with higher social capital are more supportive of corona countermeasures. From this, it can be said that it is a company that has a good relationship with employees regularly that can get the cooperation of employees in a crisis. Next, it became clear that employees with higher resilience were more supportive of corona countermeasures. It can be interpreted that the high resilience allowed him to take the current situation calmly, and to take positive action while cooperating with the company's measures against corona. Besides, it was revealed that employees who are more concerned about corona are more supportive of corona countermeasures. That is, the greater the anxiety, the stronger the desire to seek countermeasures.

We expected a negative correlation between resilience and anxiety about the corona. It was because he believed that with resilience, he would be able to control anxiety without being upset during a crisis. However, the results were reversed, and a positive correlation was found between the two. How can we interpret the unexpected positive relationship between resilience and anxiety about corona? Most of the previous studies have shown that resilience is effective in recovering from psychological shocks such as past disasters. On the other hand, the coronal disaster this time was an ongoing crisis at the time of the questionnaire. In this case, it would be more unusual to not be anxious about the corona, and it could be interpreted that those with resilience were more sensitive to the risk of corona and felt more anxious. Previous studies have also pointed out that people with higher resilience can recognize their effects more accurately in adversity and deal with them appropriately (Luthans et al., 2008; Youssef and Luthans, 2007).

However, anxiety about the corona is troublesome because it not only leads to cooperation in countermeasures against corona but also has the effect of increasing the intention to leave the job. Also, it has been clarified that the more tired the person is, the greater the anxiety about the corona, and indirectly the supportiveness of corona countermeasures. Regarding the former, it was shown in the attentional control theory (ACT; Eysenck et al., 2007) that anxious persons tend to overlook the information that is truly necessary to turn away from the information that causes anxiety. In other words, they can't calmly accept that the organization's countermeasures against coronas can calm their anxiety about coronas, and they are likely to cause thoughtless actions such as having to get out of the organization as soon as possible. On the other hand, in the latter case, in a company where employees' usual consideration for health is low, the employees who accumulated fatigue may have become more anxious due to coronal bruise, feeling a sense of crisis that further unhealthy conditions were unbearable. Therefore, they may have shown a supportive attitude toward the company's measures against corona. In previous studies, workplace fatigue was caused by physical factors such as workload, psychological factors such as work regularity and human relationships, or environmental factors such as air conditioning, sound, and lighting (Sadeghniiat-Haghighi and Yazdi, 2015). Neuroscience studies have also shown that fatigue correlates with brain health and anxiety (Kokubun et al., 2018). According to the results of this article, both a company with high social capital, a company with high employee resilience, and a company where employees feel tired or anxious were able to have employees cooperate with the company's measures against corona. Therefore, judging from the appearance, most of the companies responded well to the crisis, as are shown by the high numbers in Table 2, and it was probably difficult to notice the mechanism that occurred under the surface. However, even for employees who seemed to be cooperative with the company from the appearance, it is considered that there will be a large difference in intention to leave the job when the corona damage is over and the post-corona stage is reached. Indeed, the results in this paper show that the employees who cooperated with the company's countermeasures against corona supported by good relationships with their jobs and companies were less willing to leave. It can be said that such a company is expected to further develop with corona on its side as if turning a pinch into an opportunity, as expressed in the concrete voices of employees.

On the other hand, it seems that those who supported the corona countermeasures, supported by their resilience, also tended to lower their intention to leave. This could be interpreted as confirming that their resilience and the crisis response of the company are in the same direction and increased confidence in continuing to work at the current company. At the same time, however, it should be noted that people with high resilience tended to indirectly lower their intention to leave the job due to heightened consciousness or anxiety about the corona. As shown in Table 5, this indirect effect (-0.036) is not a large value at all, but in an organization where employees do not have a favorable impression on the work or company and it is not possible to expect a decrease in intention to leave due to this direct or indirect path, it may be something that cannot be ignored. Besides, it should be noted that people who have accumulated fatigue due to poor health management tended to increase anxiety about the corona and cooperative attitude in countermeasures against corona, and at the same time, increased the willingness to leave the job. Therefore, if anxious, employees will cooperate with the company's measures. However, their feelings will rapidly cool off after the crisis and they will leave the company, suggesting risks of using anxiety as a motivator.

**Implication**

This paper analyzed psychological data obtained from a questionnaire survey conducted on Chinese employees working for Japanese companies in China and found how employees' feelings of work/company, resilience, and even anxiety/fatigue would affect their cooperation to companies' measures against the corona and intention to leave the job. The results in this paper show that it is possible to encourage employees to cooperate with the company's measures against corona even by fueling the anxiety of corona. At the same time, however, it is also indicated that after the crisis is over, employees will face the problem of increasing intention to leave. By creating a work environment that allows employees to build good relationships with their jobs and companies, and by increasing their resilience, not only can they successfully implement crisis countermeasures, the employee's intention to leave the job can be lowered. It is important to prepare for a crisis by working to build a relationship of trust with employees and improve their resilience by this fall when there is concern about the second corona wave.

**Conclusion**

An important theme is how to maximize the cooperation of employees when dealing with crisis measures taken by the company. Therefore, to find out what kind of employees have cooperated with the company's measures in the current corona (COVID-19) crisis, and what effect the cooperation has had to these employees/companies to get hints for preparing for the next crisis, the pass analysis was carried out using awareness data obtained from a questionnaire survey conducted on 2,973 employees of Japanese companies in China. The results showed that employees with higher social capital and resilience were more supportive of the company's measures against corona and that employees who were more supportive of corona measures were less likely to leave their jobs. However, regarding fatigue and anxiety about the corona felt by employees, it was shown that it not only works to support cooperation in corona countermeasures but also enhances the turnover intention. This means that just by raising the anxiety of employees, even if a company achieves the short-term goal of having them cooperate with the company's countermeasures against corona, it may not reach the longer-term goal by making them increase their intention to leave. It is important for employees to be aware of the crisis and to fear it properly. But more than that, it should be possible for the company to help employees stay resilient, build good relationships with them, and increase their social capital to make them support crisis measurement of the company most effectively while keeping their turnover intention low.